\documentclass{article}

\usepackage{microtype}
\usepackage{graphicx}
\usepackage{subfigure}
\usepackage{booktabs} 
\usepackage{amsmath}

\DeclareMathOperator*{\argmax}{arg\,max}

\usepackage{hyperref}

\usepackage[accepted]{icml2019}

\begin{document}

\twocolumn[
\icmltitle{Grounding Object Detections With Transcriptions}






\begin{icmlauthorlist}
\icmlauthor{Yasufumi Moriya}{ireland}
\icmlauthor{Ramon Sanabria}{lti}
\icmlauthor{Florian Metze}{lti}
\icmlauthor{Gareth J. F. Jones}{ireland}
\end{icmlauthorlist}

\icmlaffiliation{lti}{Language Technologies Institute, Carnegie Mellon University, Pittsburgh, USA}

\icmlaffiliation{ireland}{Dublin City University, Dublin, Ireland}

\icmlcorrespondingauthor{Yasufumi Moriya}{yasufumi.moriya@adaptcentre.ie}
\icmlcorrespondingauthor{Ramon Sanabria}{ramons@cs.cmu.edu}
\icmlcorrespondingauthor{Florian Metze}{fmetze@cs.cmu.edu}
\icmlcorrespondingauthor{Gareth J. F. Jones}{gareth.jones@dcu.ie}

\icmlkeywords{Machine Learning, ICML}

\vskip 0.3in
]



\printAffiliationsAndNotice{\icmlEqualContribution} 

\begin{abstract}

A vast amount of audio-visual data is available on the Internet thanks to video streaming services, to which users upload their content. However, there are difficulties in exploiting available data for supervised statistical models due to the lack of labels.
Unfortunately, generating labels for such amount of data through human annotation can be expensive, time-consuming and prone to annotation errors.
In this paper, we propose a method to automatically extract entity-video frame pairs from a collection of instruction videos by using speech transcriptions and videos. 
We conduct experiments on image recognition and visual grounding tasks on the automatically constructed entity-video frame dataset of How2. The models will be evaluated on new manually annotated portion of How2 dev5 and val set and on the Flickr30k dataset.
This work constitutes a first step towards meta-algorithms capable of automatically construct task-specific training sets.

\end{abstract}

\section{Introduction}
\label{sec:intro}

\begin{figure}[h]
  \centering
    \includegraphics[width=\linewidth]{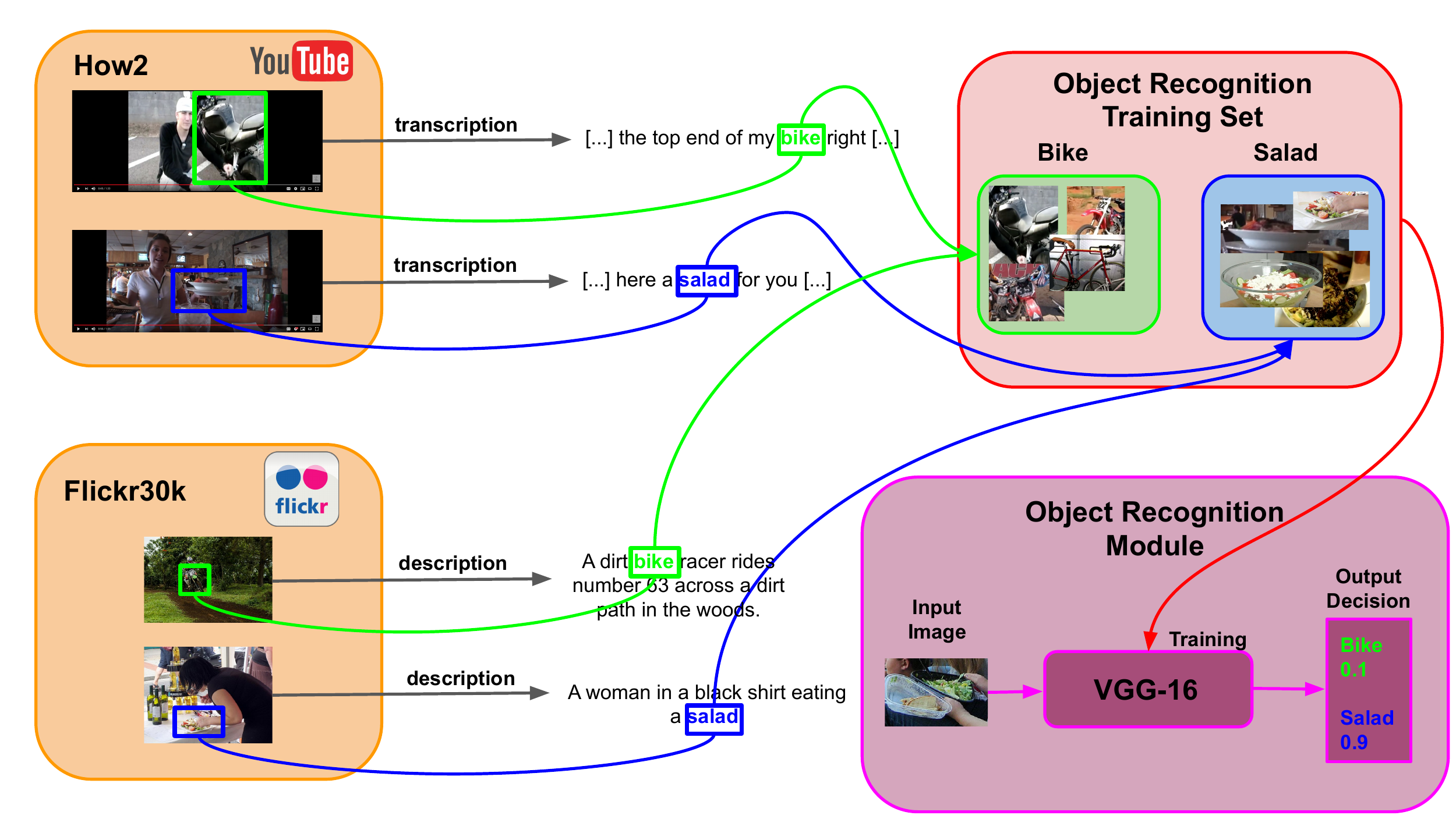}
    \caption{Our proposed approaches to automatically labelling video frames with objects based on time-aligned speech transcription. The labelled video frames can be used to fine-tune a VGG-16 object classification convolutional neural network model.}
    \label{fig:summarysystem}
\end{figure}

\begin{figure}[h]
  \centering
    \includegraphics[width=\linewidth]{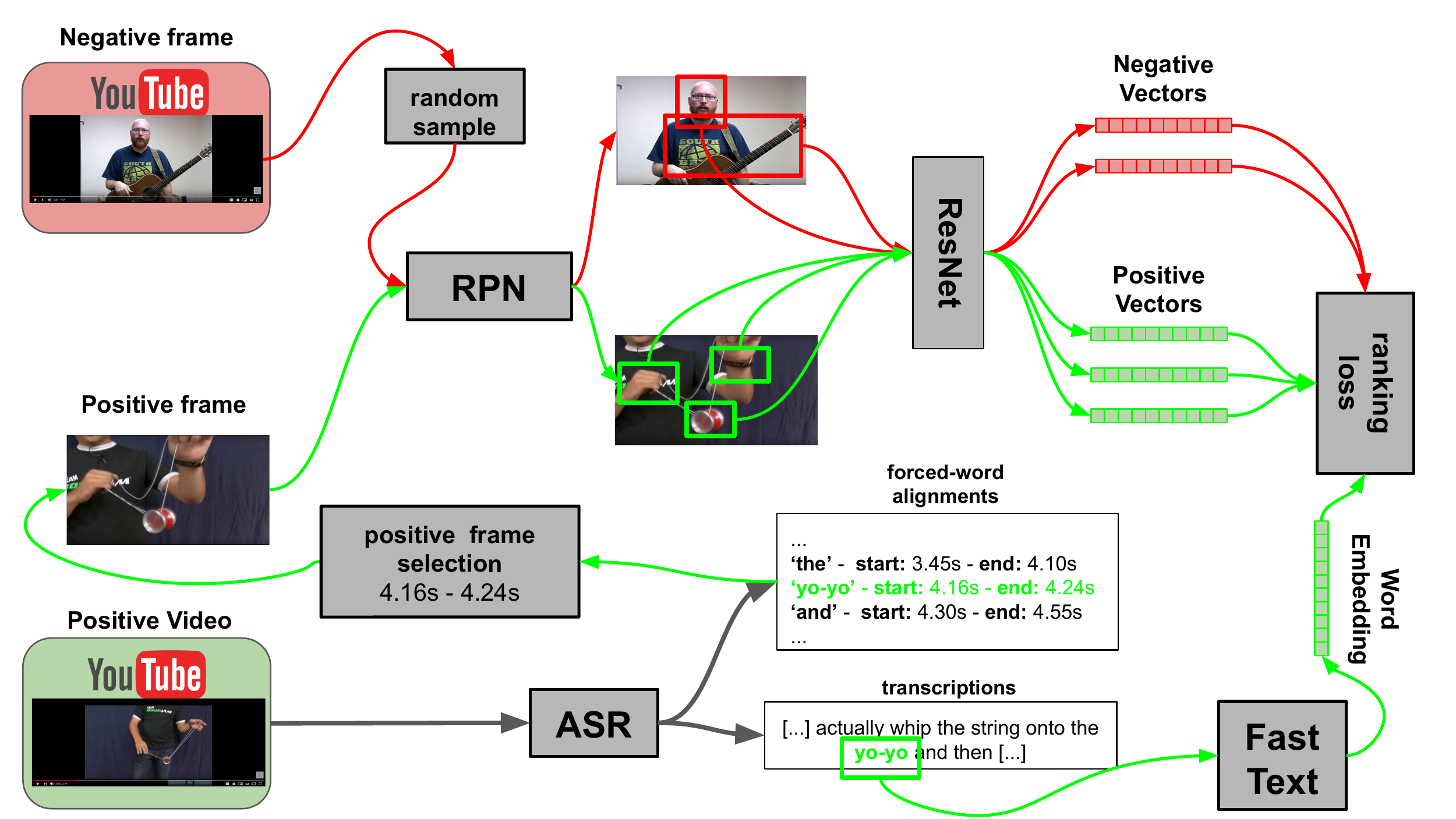}
    \caption{Our weakly-supervised visual-grounding model uses the word timestamps provided by an
    \%the 
    automatic speech recognition (ASR) forced-alignment to select a positive frame. We then sample a negative frame from a random video. A region proposal network suggests $n$ region proposals. We then extract a feature representation with a ResNet-152 model. These feature representations are later used for a ranking-loss. 
    }
    \label{fig:grounding}
\end{figure}

Audio-visual ``in the wild'' data is widely
available on the web in multiple distribution channels (\textit{e.g.} YouTube). Despite this accessibility, the lack of manual annotations for this data makes it unusable for supervised learning-based machine learning models. More specifically, neural networks require large amounts of data to estimate their parameters reliably. Crowd-sourcing platforms such as Amazon Mechanical Turk and Figure8 offer services that assign an annotation task to remote workers. However, such services are expensive, time-consuming, and quality control over annotators output is difficult. There have been previous attempts to overcome this limitation by exploiting audio-visual data for event detection systems and cooking procedures \cite{yu2014instructional,malmaud2015whats}. However, exploiting ``in the wild'' data for object detection tasks has rarely been explored, and most systems rely on a corpus that goes through partial to full manual annotation. In this work, we present object recognition and visual grounding systems that are developed from automatically generated labels (see Figure~\ref{fig:summarysystem} and \ref{fig:grounding}).



Specifically, nouns of speech transcriptions and video frames corresponding to the time-stamp of the nouns are paired.
Those video frames aligned with a noun (entity) are
used to train object recognition and visual grounding models (Section~\ref{sec:task}). Both object recognition and visual grounding are established tasks in computer vision, but the systems are rarely developed from audio-visual data. Our approach to generating labels for the tasks and summary of the dataset are described in (Section~\ref{sec:how2}). Experiments (Section~\ref{sec:exp}) on the How2 dataset~\cite{sanabria2018how2} show that it is possible to bridge speech and vision by object recognition and visual grounding.




\paragraph{Contributions} Our contributons can be summarized as follows:
\begin{itemize}
    \item We propose a novel task: grounding speech transcriptions to entities in video.
    \item We collect object-grounding annotations for the test and development set of the How2 dataset, an ``in the wild'' audio-visual corpus with speech transcriptions\footnote{The data is publically available at \url{https://github.com/srvk/how2-dataset}}. 
    \item We present a set of object recognition experiments that show how our method is able to automatically create datasets from ``in the wild'' data without the need for human intervention.
    \item Our results on the How2 dataset show that speech transcriptions can be actually grounded to visual entities. 
\end{itemize}




\section{Task overview}
\label{sec:task}


We demonstrate our automatically constructed visual-entity dataset for object recognition and weakly-supervised visual grounding tasks. Object recognition is a task to categorise a given image or a video frame into 445 classes existing in the How2 dataset. Weakly-supervised visual grounding seeks 
to draw a bounding box around
an object given an image or a video frame and a target entity.

As notation, each entity is denoted as $e_i$, where $i$ is $i$th entity. For each video frame $v_{ij}$, where $j$ is $j$th video frame of entity $e_i$, $K$ region proposals of the entity are derived. Each of the region proposals is denoted as $r_{ijk}$, where $k$ is $k$th region proposal of video frame $v_{ij}$.

\subsection{Object recognition}
\label{subsec:classification}

The use of convolutional neural networks has become a dominant approach to object recognition. For domain adaptation to
a new dataset such as How2, one of the approaches is transfer learning, where an existing model is fine-tuned on a new dataset. Specifically, the off-the-shell convolutional neural network (CNN) model is used as a feature extractor and a one layer neural network is trained to map a feature to $n$ pre-defined classes of object. Formally:
\begin{eqnarray}
    h_{ij} = f_{CNN}(v_{ij})
\end{eqnarray}
where $h_{ij}$ is a visual feature extracted from a video frame $v_{ij}$ using a convolutional neural network $f_{CNN}$. Although each video frame is paired with one entity, it is possible that one video frame shows multiple objects. To capture this effect, we experimented with both one-to-one and one-to-many training. For one-to-one training, the model is encouraged to find an entity $e_i$ for a video frame $v_{ij}$ using a softmax function: $p(e_i|v_{ij}) = Softmax(W_{cls}(h_{ij}))$ and cross entropy loss is computed given a predicted label.
For one-to-many training, the model is expected to find $m$ objects that appear in a video to which the current video frame $v_{ij}$ belongs. Therefore, instead of softmax and cross entropy loss, a sigmoid function is applied after the linear layer, and binary cross entropy loss is computed: $p(e_i|v_{ij}) = Sigmoid(W_{cls}(h_{ij}))$.

\subsection{Weakly-supervised visual grounding}
\label{subsec:grounding}

We base our visual grounding on the multiple instance learning (MIL) approach from 
\cite{huang2018-finding-it} and on the reconstruction approach from 
\cite{Rohrbach2016grounding}. This section describes these two approaches, but adapted 
for our dataset.

\subsubsection{multiple instance learning}
\label{subsec:mil}
For each video frame and given target entity, the MIL model finds an index of the region proposal most likely to correspond to the object $\bar{k}$ as follows:
\begin{eqnarray}
    \bar{k} &=& \argmax_{k} ( sigmoid( \phi ( r_{ijk} )^T \cdot \psi ( e_i )))
\label{eq:mil_out}
\end{eqnarray}
where $\phi$ is a visual encoder that transforms a cropped region of an image into visual features and $\psi$ is a text encoder that embeds a name of an entity into a dense vector. 
\begin{eqnarray}
    \phi ( r_{ijk} ) &=& W_r ( f_{CNN} ( r_{ijk} )) \\
    \label{eq:mil_vfeat}
    \psi ( e_i ) &=& W_e (f_{EMB} ( e_i))
    \label{eq:mil_wfeat}
\end{eqnarray}
A visual encoder $\phi$ consists of a convolutional neural network $f_{CNN}$ and a linear layer $W_r$. Similarly, a text encoder $\psi$ first encodes entity $e_i$ using a word embedding model $f_{EMB}$ and a linear layer $W_e$ further transforms word embedding into a textual feature. When optimising the MIL visual grounding model, only linear layers $W_r$ and $W_e$ are updated and weights of $f_{CNN}$ and $f_{EMB}$ are frozen.

The MIL visual grounding model is optimised through automatically induced entity-video frame pairs. The intuition is that any of the region proposals should be more strongly associated with the entity aligned than a randomly selected entity. Similarly, any of the entities should be more strongly connected with the video frame aligned than a randomly random video frame. To express this formally:
\begin{eqnarray}
    S_{ii} &=& \sum_{j} \max_k (\phi(r_{ijk})^T \cdot \psi (e_i))
\end{eqnarray}
and loss function $L$ is defined as follows:
\begin{eqnarray}
\begin{split}
    L = \sum^I_{i} ( &\max (0, S_{il} - S_{ii} + \delta ) +\\ &\max (0, S_{li} - S_{ii} + \delta ) )
\end{split}
\end{eqnarray}

\subsubsection{Attention and reconstruction}
\label{subsec:attention}

Another approach to detecting a bounding box of a target entity is computing attention weights given an entity and region proposals, and taking the region proposal with the highest attention weight. This is formulated as follows:
\begin{eqnarray}
    \bar{k} &=& \argmax_{k} ( f_{ATTN} ( [\phi ( r_{ijk} ); \psi ( e_i ) ]))
\end{eqnarray}
where $f_{ATTN}$ is an attention function that computes attention weights given concatenation of a visual feature of region proposal and embedded entity. Attention weight for $k$th region proposal $a_k$ is computed as follows:
\begin{eqnarray}
    \bar{a_k} &=& W_a([\phi(r_{ijk}); \psi (e_i)]) \\
    a_k &=& \frac{exp(\bar{a_k})}{\sum^N_{k=1} exp(\bar{a_k})}
\end{eqnarray}
where a linear layer $W_a$ transforms concatenation of visual and textual features, and its output is passed to the softmax function.

Optimisation of this model can be done by reconstructing an original embedded entity from visual features of region proposals to which attention weights are applied. 
\begin{eqnarray}
r_{attn} &=& W_{rec} \sum^N_{k=1}a_k\phi(r_{ijk}) \\
L_{rec} &=& \frac{1}{D}\sum^D_{d=1}(\psi(e_i)^d - r_{attn}^d)
\end{eqnarray}
where $W_{rec}$ is a linear layer to reconstruct an embedded entity from the aggregated visual features of region proposals, and $D$ is the dimensionality of embedded entities. $L_{rec}$ is essentially the mean squared error function that compares each dimensionality of the embedded entity $\psi(e_i)$ and the embedded entity reconstructed from the visual feature $r_{attn}$.

The original paper transforms textual phrases using a recurrent neural network \cite{Rohrbach2016grounding}. However, the entities extracted from the How2 dataset consist of maximum two words, and there is no need to
apply a sequential model to have textual features. For this
reason, a word embedding model is used instead of a
recurrent neural network.

\section{Dataset}
\label{sec:how2}
In this work, we use the How2 dataset to train our weakly-supervised visual grounding model \cite{sanabria2018how2}. The dataset consists of 300 hours of speech with
its corresponding transcriptions and crowd-sourced Portuguese translations. In this work, we crowd-sourced and released grounding ground-truth labels from the dev5 and val sets.

\subsection{Comparison to other visual datasets}
We reiterate that our visual grounding model is
trained on videos and speech transcriptions of the How2 dataset. Most of the existing computer vision datsets such as Microsoft COCO \cite{lin2014microsoft}, Flickr30k \cite{plummer2017flickr30k} and ReferItGame \cite{kazemzadeh2014referit} consist of images and image description, or ActivityNet \cite{heilbron2015activitynet} and YouCookII \cite{zhou2018towards} consist of videos and video description. On the other hand, How2 videos and speech transcriptions have 
not gone through any annotation process except its evaluation set. Collecting videos and speech transcriptions is relatively easier than having images or videos annotated with textual phrases. Furthermore, when no speech transcriptions are available for videos, transcriptions can be obtained using automatic speech recognition (ASR). This facilitates expansion of a set of entities that a visual grounding model learns, which
is not straightforward when any annotation process is involved in development of datasets.

\subsection{Extraction of entities}
\label{subsec:how2_extraction}
This section describes our approach to extracting entity-image pairs from How2 videos and speech transcriptions.
\begin{itemize}
    \item Force-align speech transcriptions with videos to obtain time stamp of each word uttered in videos
    \item Run Stanford Core NLP tool on speech transcripts
    to assign part-of-speech tags to words \cite{manning2014}
    \item Filter out nouns or noun phrases that are not part of ImageNet labels to retain only visible nouns \cite{russakovsky2015imagenet}
    \item Extract video frames at the end timestamp of nouns retained in the previous step  
\end{itemize}
The intuition is that when entities are uttered in speech, they are likely to appear in a visual stream. We retained entities that appear at least 5 times in 
the How2 training set. 

\subsection{Annotation of evaluation set}

In order to evaluate accuracy of our visual grounding model, we annotated video frames extracted from the dev5 and val sets of the How2 corpus with object bounding boxes using the procedures described in Section \ref{subsec:how2_extraction}. Approximately 5,000 video frames were reviewed by Amazon Mechanical Turk workers, who were asked to draw bounding boxes of a target entity. As entities from speech transcriptions are not guaranteed to appear in videos, the workers could choose the ``Nothing to label'' button. Quality of annotation was manually verified by the authors. 

\subsection{Summary of automatically constructed dataset}

This section gives a summary of the automatically constructed dataset. Overall, there are 139,867 video frames extracted from the training set, and 2,301 video frames from the dev5 and val sets. The number of entities 
labelled was
533, when there is no distinction between singular and plural, and 445, when there is the distinction. The numbers reported from now on assume that singular and plural variants are merged into the same class. The 10 most frequent entities in train and dev5+val are summarised in Table \ref{table_entities}. As can be seen in the table, frequent entities refer to a human body part both in train and dev5+val set. Apart from body parts, the dataset includes ``ball'' (3,306 times in the training set and 30 times in the combined dev5+val set), ``horse'' (429 times in train and 22 times in dev5+val), and ``knife'' (327 times in 7 times in dev5+val), since the instruction videos demonstrate a variety of activities such as sports, animal cares, and cooking.

Initially, 5,267 video frames from the dev5 and val sets were prepared to go through human annotation with a target entity label. However, the quality of annotation was not satisfactory on some of the video frames for
the following reasons: (1) a video frame contains a target entity, but bounding boxes are not correctly drawn or ``Nothing to label'' was selected, and (2) a video frame does not contain a target entity, but ``Nothing to label'' was not selected and bounding boxes were drawn. By excluding the rejected work, 4,775 video frames remained, where 2,331 were ``Nothing to label'' and 2,444 were annotated. Since the dev5 and test sets retained some entities that do not appear in the training set, the total number of video frames for evaluation was
2,301.
It is reported that target entities in YouCook validation set appear in 51.32\% of the total frames \cite{shi2019not}. 
As compared to this, our automatic data construction approach collected 51.12\% of positive labels, which 
is comparable to the quality of the YouCook videos whose video description was manually created for a
visual grounding task.

\begin{table}[th]
\centering
\caption{Top 10 most frequent entities in train and dev5+val}
\bigskip
\begin{tabular}{lllll}
\toprule
        & train & frequency & dev5+val  & frequency   \\ \midrule
top1    & hand   & 5,576    & hand       & 131 \\
top2    & foot   & 4,224    & foot       & 69  \\
top3    & people & 3,420    & hair       & 68  \\
top4    & back   & 3,334    & body       & 64  \\
top5    & ball   & 3,306    & arm        & 58  \\
top6    & leg    & 3,026    & leg        & 58  \\
top7    & water  & 2,595    & people     & 56  \\
top8    & body   & 2,579    & shoulder   & 38  \\
top9    & top    & 2,455    & skin       & 38  \\
top10   & line   & 2,430    & face       & 36  \\ \bottomrule
\end{tabular}
\label{table_entities}
\end{table}

\section{Related work}
\label{sec:relwork}



\paragraph{Semi-supervised Classification}

Supervised object detection and classification models have been 
studied in many previous works~\cite{Ren2015faster,He2017mask, Redmon2016you,Redmon2017YOLO9000}. However, these approaches, 
rely on large
annotated datasets that are usually expensive to create and prone to error. For this reason, to overcome this difficulty, there has been much
work on weakly-supervised models for object and event classification~\cite{zhuang2017attend, guo2018curriculumnet, yu2014instructional, chang2015searching}. More concretely,~\cite{zhuang2017attend, guo2018curriculumnet} exploited crawled images from
the web to construct large-scale image classification datasets without the need of any human intervention. Their methods are based on reducing the negative impact that web labels could have. More related to this work, \cite{yu2014instructional, chang2015searching} used the alignments between ``in the wild'' collected videos and their 
transcripts to localize events and use them for training a multimodal event detection algorithm.

\paragraph{Grounding} 

Multimodal retrieval, also, arguably, called grounding, has been studied in many previous works in
the language processing community~\cite{harwath2018vision,kamper2017visually, arandjelovic2018objects, aytar2017see}. More concretely, \cite{harwath2018vision,kamper2017visually} use images as anchor modality to retrieve a textual representation from a speech query. On the other hand, \cite{arandjelovic2018objects, aytar2017see} propose a more general model retrieval approach where queries can be done across multiple modalities.

From a more specific grounding perspective, \citet{malmaud2015whats} propose an approach to align cooking recipes with ASR transcripts using a hidden Markov model and a keyword spotting system. In the image domain, \citet{karpathy2015deep} generate natural language descriptions of regions by aligning image bounding boxes to its description with a neural approach. More recently, \cite{Rohrbach2016grounding, harwath2018jointly} ground images to their
textual and acoustic (\textit{i.e.} speech) descriptions respectively. Our grounding approach (Section~\ref{subsec:grounding}) is inspired by \cite{Rohrbach2016grounding}. \cite{shi2019not,zhou2019grounded, zhou2018weakly, huang2018-finding-it,zhao2018sound,ephrat2018looking} propose approaches to
grounding speech and textual representations to videos. More concretely, ~\cite{zhao2018sound,ephrat2018looking} ground sound to image regions and use it~\textit{a posteriori} to perform
source sound localitzation and isolation. On the other hand, \cite{shi2019not, zhou2019grounded, zhou2018weakly, huang2018-finding-it} ground textual video descriptions to video. All previously cited work uses cleaned controlled, human labeled datasets such as YouCookII~\cite{huang2018-finding-it}, RoboWatch~\cite{sener2015unsupervised} or activtiyNet~\cite{zhou2019grounded}, as opposed to our
work where we use How2, an ``in the wild'' collected dataset.

In parallel with this work, \citet{amrani2019toward} propose a similar approach to
automatically create 
object recognition datasets with How2. There are three fundamental differences between \cite{amrani2019toward} and our work. First, \citet{amrani2019toward} evaluate their approach on a newly defined test set that contains videos from the training set. Our experiments use the official data partitions of the How2 dataset~\cite{sanabria2018how2}. Second, their approach uses up to 11 categories as opposed to our work, where we use up to 533 classes (445 when singular and plural entities are merged). Finally, \citet{amrani2019toward} limit their experimentation results to the How2 dataset, as opposed to our work where we analyze the generalizability of our approach by also reporting results on image descriptions.

\section{Experiments}
\label{sec:exp}

\subsection{Flickr30k}
\label{subsec:flickr}

To evaluate generalisability of object recognition and visual grounding models developed from our visual-entity How2 corpus, we adopted the Flickr30k dataset for another evaluation corpus \cite{plummer2017flickr30k}. The dataset consists of more than 30,000 images, from which 1,000 images each are assigned to validation and test sets. For each image, there are 5 image descriptions created by Amazon Mechanical Turk workers. Furthermore, the dataset is annotated with bounding boxes that capture phrases of image descriptions (e.g., ``A stop sign''). For evaluation, we only use the validation and test sets of the dataset, except that we fine-tuned a VGG16 for object recognition on training data to compare to models trained on How2.

For object recognition, each phrase of an image description is compared against the How2 visual entities. When there is a common entity between a phrase and the set of How2 entities, an Flickr30k image is labelled with the entity. When there are multiple entities that exist in the How2 entities, we take the last word of the phrase (e.g., ``corner'' from ``A street corner''). In the Flickr30k validation set, 6,927 phrases contain a single entity common with the How2 entities and 321 phrases multiple entities with How2. In the Flickr30k test set, 6,868 phrases contain a single entity common with the How2 entities and 370 phrases multiple entities with How2. None of the phrases in the 10 images in the validation and test sets share common entities with the How2 dataset. For visual grounding, bounding box labels are extracted using the same approach, 
some phrases are not annotated with bounding boxes, because not all of the phrases can be captured by a bounding box. There are 3,188 bounding boxes for the test set and 3,232 bounding boxes for the validation set.

\subsection{Experimental setup}
\label{subsec:exp_setup}

For object recognition, we use the VGG16 model as a feature extractor. While this is not a state-of-the-art model, it has shown reasonable accuracy on the task, and fine-tuning can be accomplished more quickly
than other models because of the number of parameters that the model has. The linear classification layer is set to transform input image features of 4,096 dimensions to 256 dimensions. 
The feature then goes through the ReLU layer and the dropout layer with 0.5 dropout rate. Finally, the last linear layer transforms the input to the 445 
unique entities in the How2 visual-entity corpus. The mini-batch size was set to 64, with 
learning rate set to
0.0001. The Adam optimizer was used to update the weights of the classification layers. The model was trained for 20 epochs. 

For visual grounding, 20 candidate object bounding boxes were extracted from both the How2 and Flickr30k datasets using a region proposal network belonging to Mask-RCNN provided by Facebook Research \cite{massa2018mrcnn}. We chose the region proposal network of the Mask-RCNN with 
ResNeXt101 as the backbone. This model was pre-trained on the COCO 2017 dataset. For each candidate bounding box, the ResNet 152 model pre-trained on the ImageNet dataset, extracts 2048 dimensional visual features. The word embedding model was trained on a speech transcript of the training set of the How2 dataset using fastText model. The dimensionality
of word embedding was set to 100. 

The MIL model consists of one visual feature encoder and one word embedding encoder. The visual feature encoder has one fully connected layer that transforms 2048 dimensional features into 512, ReLU and dropout with rate 0.2. An additional fully connected layer is then applied to the visual features and the final dimensionality of the features is 100. The word embedding encoder is one fully connected layer that transforms 100 dimensioal vectors to 100 dimensional vectors. The model was trained for 20 epochs with a learning rate of 0.00001 using the Adam optimizer, and delta was set to 0.01. We observed that higher a learning rate easily leads the model to being stuck in local minima. 

The reconstruction model consists of one visual feature encoder and one word embedding encoder. The visual feature encoder transforms 2,048 dimensional vectors into 100 dimensional vectors, and the word embedding encoder encodes 100 dimensional input to 100 dimensional output. Those features are then concatenated and go through the attention layer. The model was trained for 20 epochs with a 
learning rate 
set to 0.001 using the Adam optimizer. 

\subsection{Object recognition results}
\label{subsec:res_cls}

\begin{table}[t]
\centering
\caption{Results of image classification models on dev5+val set of How2. Results are reported for
top $k$ accuracy. A `single'' model was fine-tuned with softmax targeting at one label for each example, 
and a ``multi'' model with sigmoid targeting at multiple labels for each example.
``flickr30k'' is a model fine-tuned on image-label pairs for comparison to the other models.} 
\bigskip
\begin{tabular}{lllll}
\toprule
                & top1 & top5  & top10 & top20 \\ 
\midrule
single          & 9.95 & 25.27 & 36.47 & 49.44 \\
multi           & 8.81 & 23.69 & 34.81 & 49.11 \\
flickr30k       & 0.76 & 7.00  & 11.73 & 18.64 \\
\bottomrule
\end{tabular}
\label{table_cls_how2}
\end{table}

Table \ref{table_cls_how2} summarises results of image classification on the How2 dev5+val set. As described in Section \ref{subsec:classification}, the VGG16 model was fine-tuned using two different training criterion. An additional model was fine-tuned on Flickr30k data using multi label training, 
since each Flickr image contains multiple entities in its description. The results are reported in terms of the accuracy of the top $k$ predictions from the model, since
each video frame is annotated with one object on the How2 dev+val setS.

As shown 
in Table \ref{table_cls_how2}, the model trained on using the softmax funciton with a single label for
one example showed better accuracy than the model 
trained using a sigmoid function with multiple labels. It is understandable that accuracy at top 1 prediction is just less than 10\%, as any instruction video captures multiple objects (e.g., a speaker figure and various body parts). In fact, when the top 5 predictions of the ``single'' model are considered at one time, accuracy increases to 25\%, and further to just less than 50\%, when taking top 20. It also turned out that the model fine-tuned on Flickr data had difficulties recognising objects on How2 dataset, even though te labels of the Flickr dataset are derived from image descriptions created by human annotators.

\begin{table}[th]
\centering
\caption{Results of object recognition models on validation and test sets of the Flickr30k dataset. Results are reported for 
mean average precision (MAP) of top k recognised entities.} 
\bigskip
\begin{tabular}{llll}
\toprule
    &   MAP@5 & MAP@10 & MAP@20 \\
\midrule
\textbf{validation} \\
single          & 6.47  & 7.48   & 8.34   \\
multi           & 7.86  & 8.74   & 9.54   \\
flickr30k       &37.11  &41.74   &44.34   \\
\midrule
\textbf{test}\\
single          & 6.45 & 7.53 & 8.37 \\
multi           & 7.22 & 8.25 & 8.98 \\
flickr30k       &35.69 &39.91 & 42.5 \\
\bottomrule
\end{tabular}
\label{table_cls_flick}
\end{table}

Table \ref{table_cls_flick} shows the object recognition results on the Flickr30k dataset. Since each image on Flickr30k can have multiple labels, the models are evaluated through mean average precision (MAP) of the top $k$ predictions. It turned out that using ``multi'' label training led to better generalisation in this case.
The ``multi'' system consistently shows better MAP than the system trained using the softmax function.
As expected, the model fine-tuned on Flickr30k labels performed much better than models on How2. 

\begin{figure}[h]
  \centering
    \includegraphics[width=\linewidth]{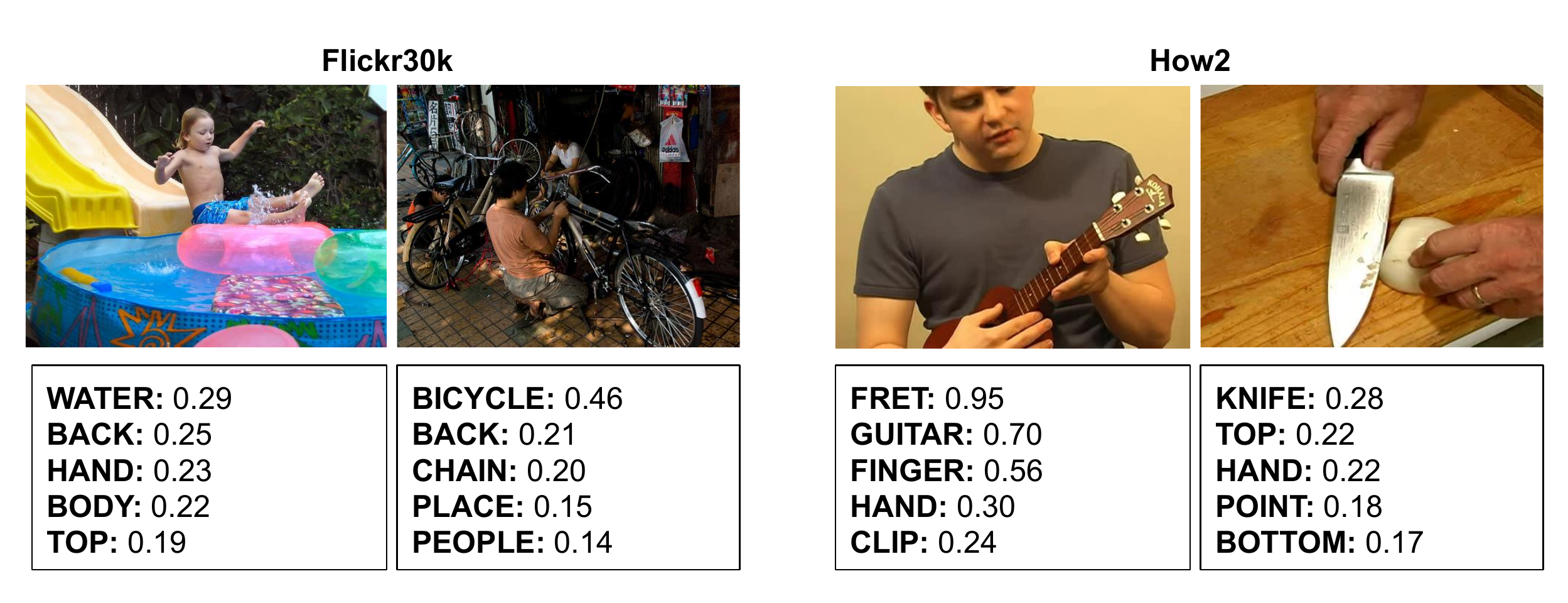}
    \caption{Output examples of object recognition produced by the 
    VGG-16 model trained on our automatically generated training set.} 
    \label{fig:recognition_output}
\end{figure}

Figure \ref{fig:recognition_output} shows example 
outputs of the ``multi'' system fine-tuned on How2. As can be seen from
the figure, objects that are not part of popular classes in the How2 training set were predicted with good probabilities. For example, ``knife'' occurred 327 times and ``fret'' 399 times. This shows that automatic construction of a visual dataset using speech transcription and videos offers reliable labels for object recognition.

\subsection{Visual grounding results}
\label{subsec:res_grounding}

\begin{table}[th]
\centering
\caption{Results of weakly-supervised models on the dev5 and test sets of he How2 dataset. The upper-bound is the best score that the models can achieve given candidate object bounding boxes.}
\bigskip
\begin{tabular}{llll}
\toprule
                    & IoU 0.5 & IoU 0.3 & IoU 0.1 \\
\midrule
\textbf{upperbound} & 43.6    & 58.8    & 82.4   \\
random              & 7.6     & 18.0    & 40.6   \\
reconstruction      & 7.9     & 19.2    & 40.2   \\
MIL                 & 9.3     & 22.6    & 49.3   \\
\bottomrule
\end{tabular}
\label{table_grounding_how2}
\end{table}

Table \ref{table_grounding_how2} summarises results of visual grounding systems trained on the How2 dataset. The results are reported in terms of Intersection over Union (IoU) between a predicted candidate region and a gold standard region. When a predicted candidate region had an overlap with any of the gold standard regions over the threshold, the prediction was regarded as positive. Due to the constraints on region proposals, the best accuracy achievable for a certain threshold is shown
in the table. 

As can be seen in Table \ref{table_grounding_how2}, the random baseline turned out to be the worst, with
reconstruction and MIL approaches being
slightly better than the random baseline. MIL consistently produced the highest accuracy on any threshold settings. 

Table \ref{table_grounding_flickr} shows results of our visual grounding systems applied to Flickr30k data. In general, a higher upper-bound was obtained for Flickr30k data. This is possibly because the region proposal network was trained on the COCO 2017 data, and the domain of this data is similar to that of Flickr30k. The results show that both reconstruction and MIL approaches produced better accuracy than the random baseline. Furthermore, the MIL system was consistently the best for
any threshold.
This result was also observed for the How@ dataset. Overall, even though the approaches used were not state-of-the-art, the models could learn visual-textual relationships from labels automatically generated from the How2 dataset. Further improvement can be expected by adopting a more sophisticated visual grounding approach or introducing removal of false alarm video frames. 

\begin{table}[th]
\centering
\caption{Results of weakly-supervised models on validation and test set of Flickr30k. The upper-bound is the best score that the models can achieve given the candidate object bounding boxes.}
\bigskip
\begin{tabular}{llll}
\toprule
                    & IoU 0.5 & IoU 0.3 & IoU 0.1 \\
\midrule
\textbf{validation} \\
\textbf{upperbound} &  65.6  &  76.1   & 89.8    \\
random              &  15.6  &  27.2   & 47.0    \\
reconstruction      &  18.4  &  31.8   & 54.1    \\
MIL                 &  18.9  &  35.6   & 58.5    \\
\midrule
\textbf{test} \\
\textbf{upperbound} &  65.7  &  75.9   & 90.0    \\
random              &  15.4  &  28.6   & 45.6    \\
reconstruction      &  18.9  &  30.7   & 51.6    \\
MIL                 &  19.6  &  36.4   & 58.1    \\
\bottomrule
\end{tabular}
\label{table_grounding_flickr}
\end{table}

\section{Conclusion}
In this paper, we have investigated the use of audio-visual data ``in the wild'' to induce image labels for object recognition and visual grounding tasks. Overall, we have collected over 500 types of labels from 300 hours of video adata. It has been
demonstrated that simple approaches can
learn relationships between vision and entities derived from speech transcription. This can be an advantage of removing label annotation from visual data construction. Future work will include refinement of visual grounding approaches to improve accuracy of the task on the How2 dataset.


\section{Acknowledgement}

This work was partially supported by Science Foundation Ireland as part of the ADAPT Centre (Grant 13\//RC\//2106) at Dublin City University.

\nocite{langley00}

\bibliography{main}
\bibliographystyle{icml2019}



\end{document}